\documentclass[12pt,preprint]{aastex}

\shorttitle{Errors in ZEUS}
\shortauthors{Falle}

\begin{document}

\title{Rarefaction Shocks,  Shock Errors and Low Order  of Accuracy in
ZEUS}


\author{S. A. E. G.  Falle}
\affil{Department of  Applied Mathematics, University  of Leeds, Leeds
LS2 9JT, UK}

\begin{abstract}
We show that  there are simple one dimensional  problems for which the
MHD   code,  ZEUS,  generates   significant  errors,   whereas  upwind
conservative schemes perform very well on these problems.
\end{abstract}




\section{Introduction}

ZEUS  is a  freely  available MHD  code  that is  widely  used by  the
Astrophysical  community.   Although Stone  \&  Norman (1992a,b)  give
results for  the Sod  problem (Sod 1978)  and its MHD  equivalent, the
Brio and  Wu problem (Brio \& Wu  1988), ZEUS does not  appear to have
been  tested  on  a wide  range  of  Riemann  problems such  as  those
described in e.g.  Dai \& Woodward (1994), Ryu \& Jones (1995), Falle,
Komissarov \& Joarder (1998)and  Balsara (1998).

Since  ZEUS  is  neither  upwind  for all  characteristic  fields  nor
conservative, we  might expect it  to perform significantly  less well
than upwind conservative codes (e.g.  Brio \& Wu 1988; Dai \& Woodward
1994; Ryu  \& Jones 1995;  Falle, Komissarov \& Joarder  1998; Balsara
1998; Powell  et al. 1999).  As we  shall see, this is  indeed true in
the sense  that there are  a number of  simple problems for  which the
ZEUS solution contains significant errors that are absent in solutions
calculated with an upwind conservative scheme.

\section{Rarefaction Shocks}

\begin{figure}[h]
\plotone{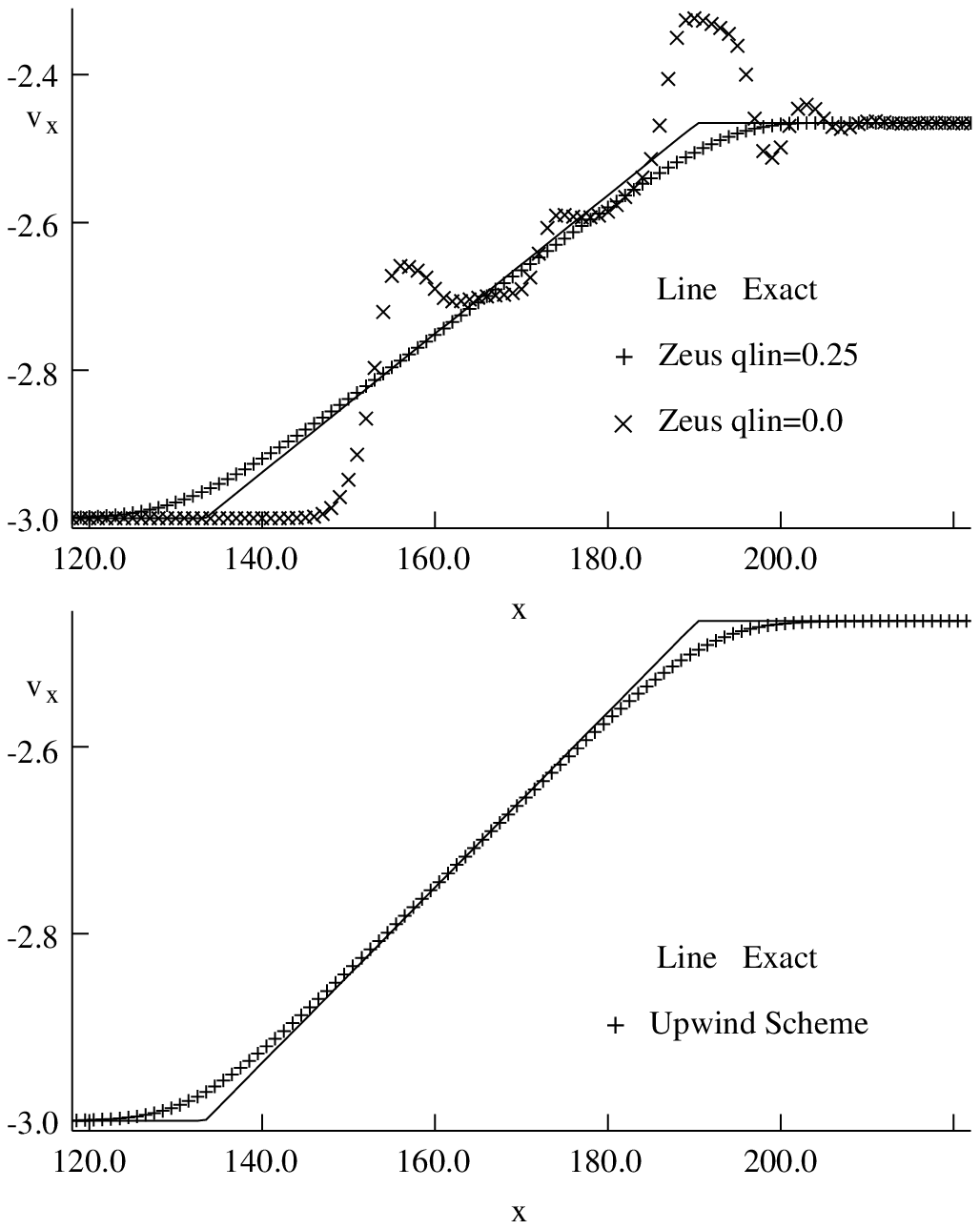}
\caption{Gas rarefaction at  $t = 80$. Left state:  \mbox{$\rho = 1$},
\mbox{$p_g  = 10$},  \mbox{$v_x =  -3$}.  Right  state:  \mbox{$\rho =
0.87469$},   \mbox{$p_g   =   8$},   \mbox{$v_x  =   -2.46537$}.   The
discontinuity is at $x = 700$ at $t = 0$ and $\Delta x = 1.0$.}
\end{figure}

\begin{figure}[h]
\plotone{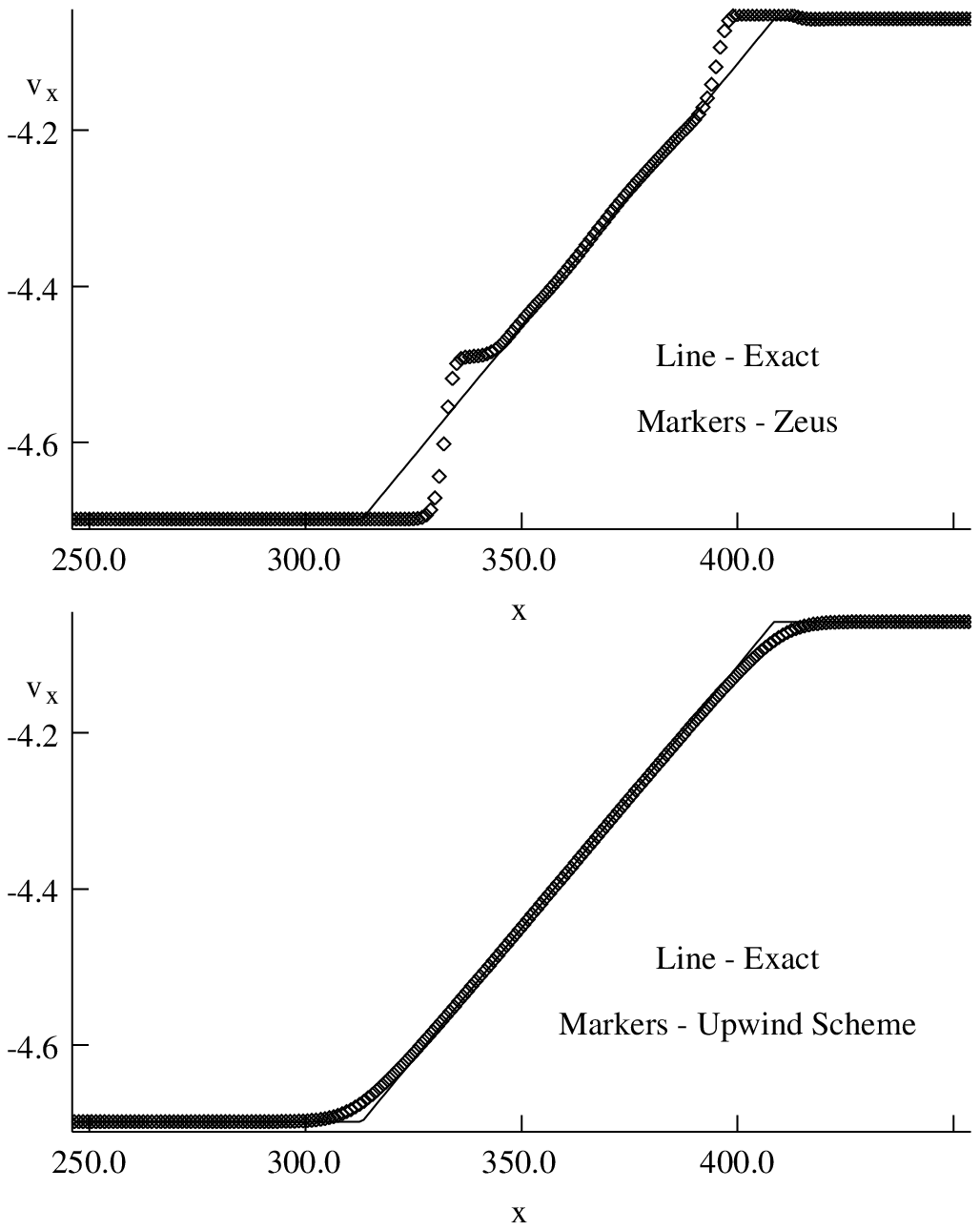}
\caption{Fast rarefaction at $t = 100$. Left state: \mbox{$\rho = 1$},
\mbox{$p_g   =  0.2327$},   \mbox{$v_x  =   -4.6985$},   \mbox{$v_y  =
-1.085146$}, \mbox{$B_x = -0.7$}, \mbox{$B_y = 1.9680$}.  Right state:
\mbox{$\rho = 0.7270$}, \mbox{$p_g = 0.1368$}, \mbox{$v_x = -4.0577$},
\mbox{$v_y  = -0.8349$},  \mbox{$B_x =  -0.7$}, \mbox{$B_y  = 1.355$}.
The discontinuity  is at $x =1000$  at $t =  0$ and $\Delta x  = 1.0$.}

\end{figure}

\begin{figure}[h]
\plotone{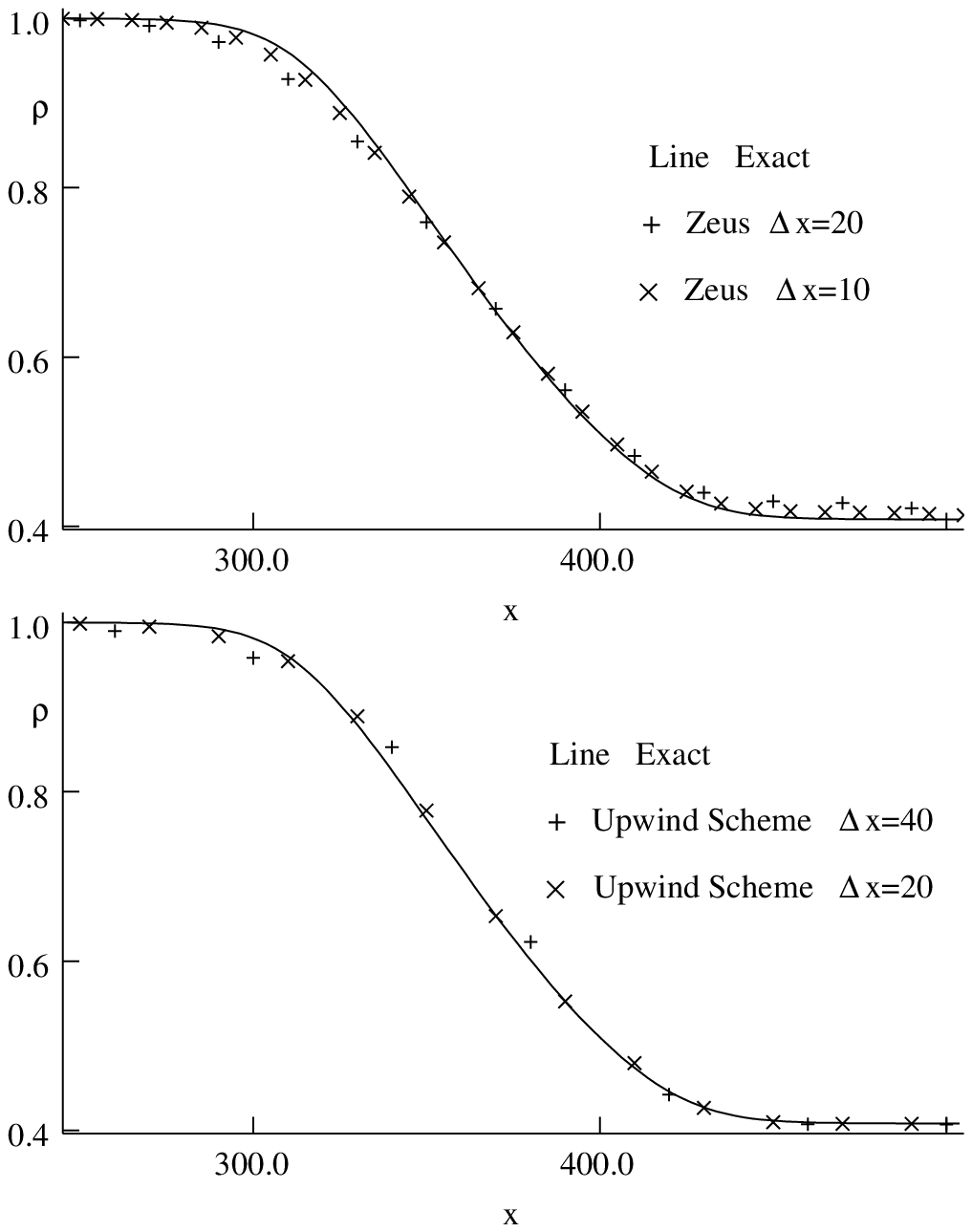}
\caption{Smooth gas rarefaction  at $t = 50$. At  $t = 0$,\mbox{$v_x =
0.5  \left[ {1  +  tanh \{  0.1(x  - 400)  \} }\right]$},  \mbox{$\rho
\rightarrow  1$}, \mbox{$p_g \rightarrow  1$} as  \mbox{$x \rightarrow
-\infty$}. The ZEUS calculation is with $\mbox{qlin} = 0.25$.}
\end{figure}

\begin{figure}[h]
\plotone{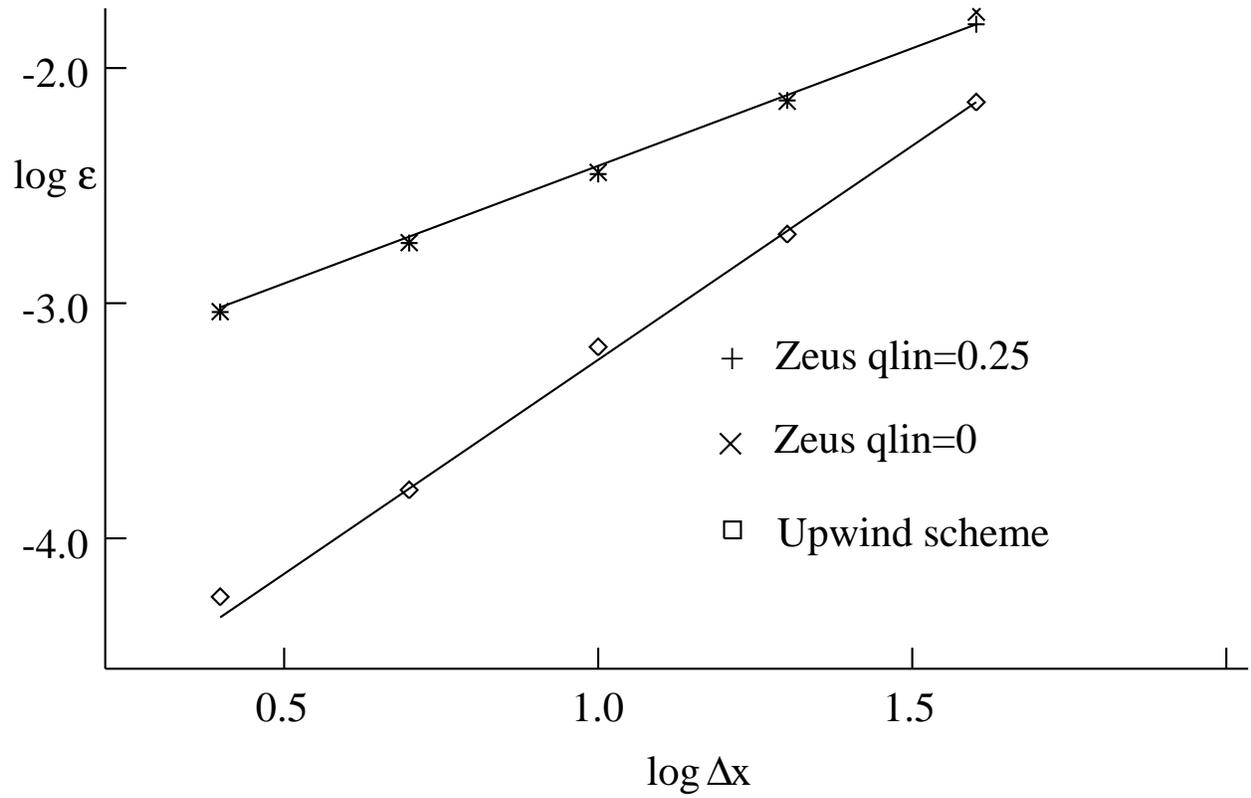}
\caption{Convergence rates: $\epsilon$ is  the $L_1$ norm of the error
in the  density for  the smooth  rarefaction in $200  \le x  \le 700$.
Lines with slopes $1$ and $2$ are also shown.}
\end{figure}

\begin{figure}[h]
\plotone{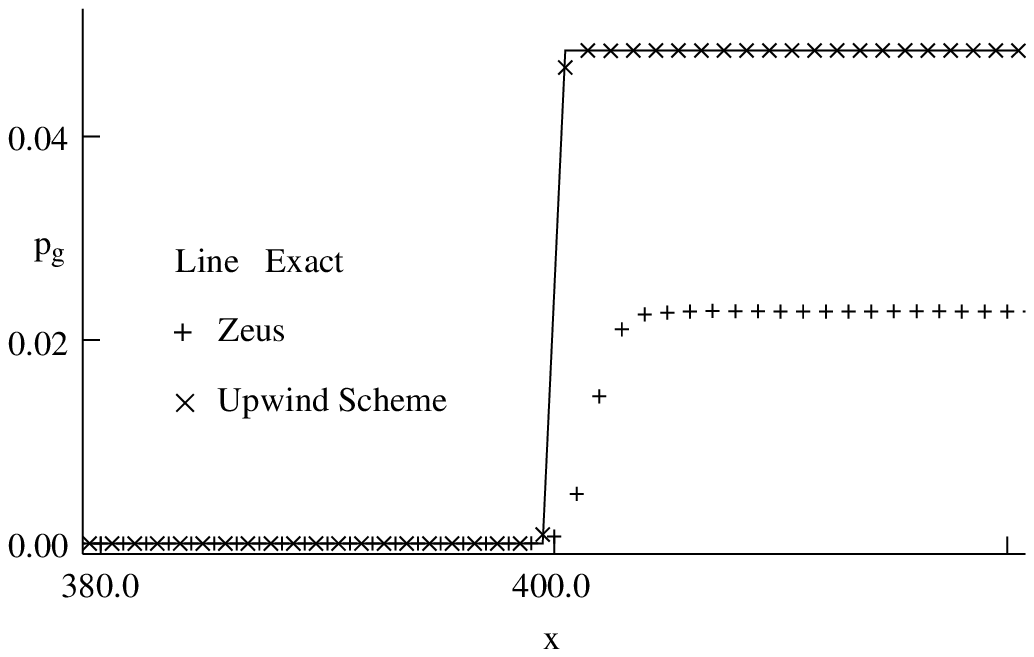}
\caption{Gas   pressure  in  stationary   fast  shock.    Left  state:
\mbox{$\rho  =  1$},  \mbox{$p_g  =  10^{-6}$},  \mbox{$v_x  =  1.5$},
\mbox{$v_y  = 0.0$}, \mbox{$B_x  = 0.1$},  \mbox{$B_y =  1.0$}.  Right
state:  \mbox{$\rho =  1.6111$}, \mbox{$p_g=  0.04847$},  \mbox{$v_x =
0.9310$},  \mbox{$v_y =  0.04104$}, \mbox{$B_x  = 0.1$},  \mbox{$B_y =
1.6156$}.}
\end{figure}

\begin{figure}[h]
\plotone{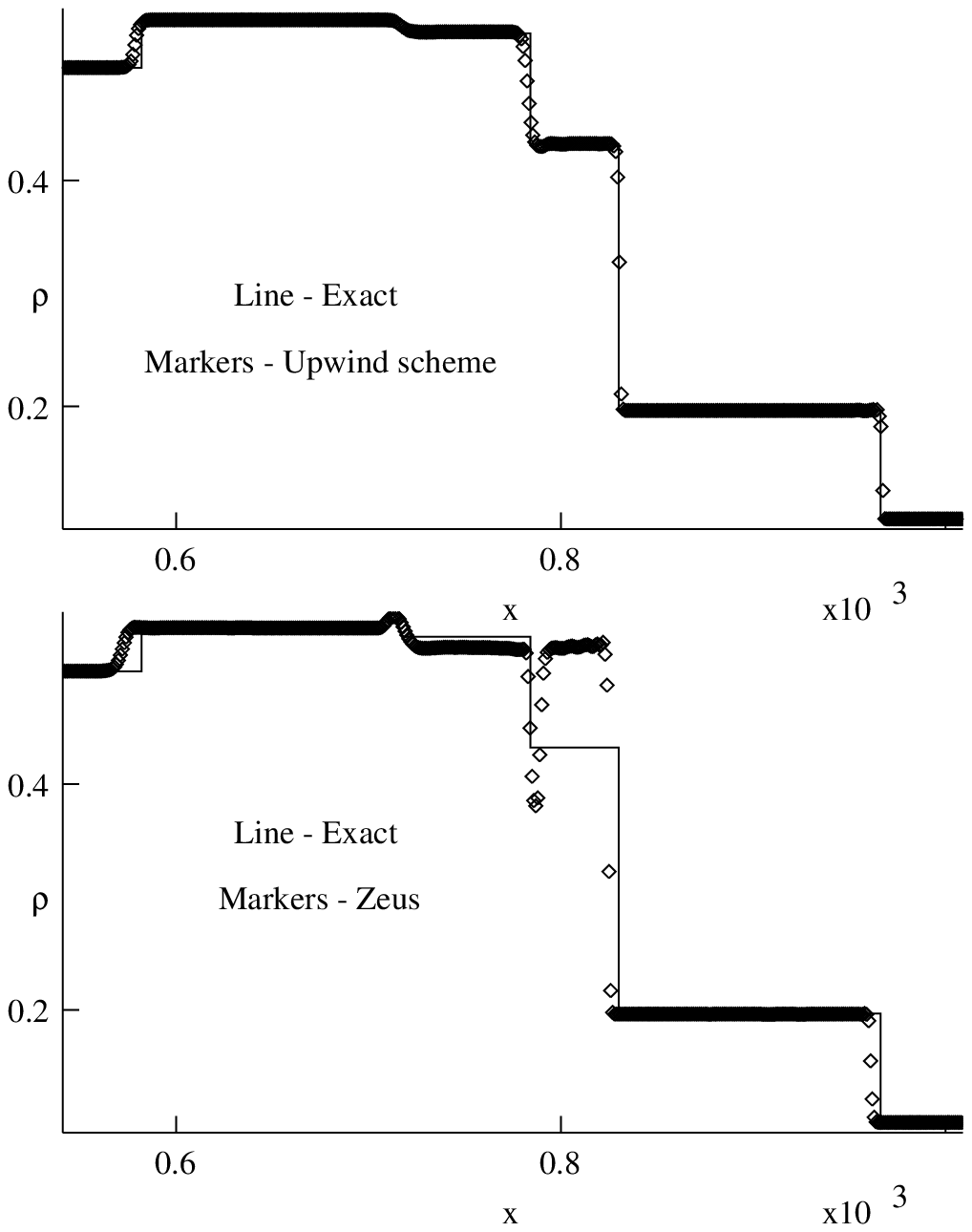}
\caption{Density in a Riemann problem at \mbox{$t = 30$}.  Left state:
\mbox{$\rho = 0.5$}, \mbox{$p_g = 10$}, \mbox{$v_x = 0$}, \mbox{$v_y =
2$}, \mbox{$B_x = 2$}, \mbox{$B_y = 2.5$}.  Right state: \mbox{$\rho =
0.1$},  \mbox{$p_g=  0.1$},  \mbox{$v_x  = -10$},  \mbox{$v_y  =  0$},
\mbox{$B_x = 2$}, \mbox{$B_y =  2$}. The exact solution was calculated
using the  Riemann solver described  in Falle, Komissarov  and Joarder
(1998)}
\end{figure}

Figures 1 and  2 show that ZEUS generates  rarefaction shocks for both
pure gas rarefactions and  fast magnetosonic rarefactions, whereas the
upwind conservative  scheme described in Falle,  Komissarov \& Joarder
(1998) gives  quite satisfactory  results.   In both  cases, the  ZEUS
solutions  are sensitive  to the  inertial frame  and  the rarefaction
shocks can be removed by  a Galilean transformation that increases the
x velocity sufficiently.

These rarefaction  shocks are steady  structures whose width  does not
increase  with time. Since  the effect  of the  nonlinear terms  is to
spread such structures, it is clear that the truncation errors in ZEUS
must be  anti-diffusive in these  cases. The most  obvious explanation
for this  is that ZEUS  is second order  in space, but first  order in
time since this  can lead to an anti-diffusive  term in the truncation
error.  For example, the upwind scheme can be made first order in time
and second order in space by omitting the preliminary first order step
and  in that  case  it can  be  shown to  be  anti-diffusive and  also
produces rarefaction shocks.

Although ZEUS is second order  in space and time for linear advection,
the use  of a partially updated  velocity in the  advection step means
that  it  is  only  first  order  in  time  if  the  velocity  is  not
constant. Further  evidence that this is  the cause of  the problem is
provided by the sensitivity of  the rarefaction shocks to the Galilean
frame  and the fact  that they  disappear when  the Courant  number is
reduced from  $0.5$ to  $0.1$, whereas they  become much worse  if the
Courant number is increased above $0.5$.

ZEUS has  a facility  for adding a  linear artificial  viscosity whose
magnitude is determined by the parameter $qlin$.  The addition of such
a viscosity removes the anti-diffusive terms by reducing the scheme to
first order in space for  everything except linear advection.  For the
gas  rarefaction, $qlin  = 0.25$  cures the  problem and  seems  to be
optimal for a  global Courant number of $0.5$, but it  is too large if
the local Courant number associated with the wave is small.  Since the
linear viscous  term must balance  an anti-diffusive term  that scales
like the  timestep, it  would be  better if the  viscous term  that is
implemented  in  ZEUS were  multiplied  by  the  local Courant  number
associated  with  the  wave   that  is  causing  the  problem.   Since
rarefaction  shocks only  arise for  rarefactions associated  with the
sound wave  with the  largest speed  relative to the  grid, it  is the
smallest local Courant number that is appropriate.

In MHD  the situation  is even worse  since, although  the rarefaction
shocks in  the fast  rarefaction can be  removed by  setting $qlin=1$,
this makes the scheme very diffusive for other waves. Furthermore, the
required value of $qlin$ depends  on the particular problem.  It might
be  possible to  avoid  such a  large  value of  $qlin$  by adding  an
appropriate artificial  resistivity, but the code has  no facility for
this.

Figures 3  and 4 show that,  even for an  initially smooth rarefaction
wave, ZEUS is  significantly less accurate than an  upwind scheme. The
results are  for $qlin=0.25$,  but Figure 4  shows that ZEUS  is still
first order even without this.  In contrast, it is evident from Figure
4 that the  rate of convergence of the upwind  scheme is second order.
Note  that the  upwind  scheme  also has  an  artificial viscosity  as
described in  Falle, Komissarov \&  Joarder (1998), but since  this is
applied in the Riemann solver, it  does not reduce the order in smooth
regions. Incidentally, ZEUS  performs even worse if one  does not take
the  staggered  grid  into  account  in setting  up  the  the  initial
solution.   Furthermore, for both  codes, point  samples were  used to
project  the exact  solution onto  the grid,  which is  reasonable for
ZEUS, but is somewhat unfair to a conservative scheme.

The  upwind   scheme  produces   reasonable  results  at   the  lowest
resolution,  even though  this corresponds  to only  $2$ cells  in the
rarefaction at the initial time, whereas ZEUS needs at least $4$ cells
for the same accuracy. For a three dimensional calculation, this would
require $24$ times  the computing time and $8$  times the memory since
ZEUS  is  about $2/3$  times  the speed  of  the  upwind scheme.   The
disparity in efficiency is actually greater than this because for both
codes the  Courant number was set  to the ZEUS default  value of $0.5$
for all  cases described  in this  paper. The upwind  code can  run at
larger Courant numbers than this,  whereas even $0.5$ can be too large
for ZEUS for some Riemann problems.  Of course, the slower convergence
of ZEUS also  means that the situation would be  even worse if greater
accuracy were required.

\section{Shock Errors}

Since ZEUS  is not  conservative, we expect  it to generate  errors at
shocks which  cannot be reduced  by increasing the resolution.   As it
turns out,  these errors are small ($  < 5 \%$) for  pure gas dynamics
and are entirely absent for  an isothermal equation of state. However,
they can be significant for adiabatic MHD .

Figure  5 shows  that,  for  a nearly  perpendicular  fast shock,  the
post-shock gas pressure in the ZEUS solution is too low by a factor of
2.  In  contrast, the  conservative  upwind scheme  gets the  solution
exact to  rounding. It is  true that this  is a somewhat  extreme case
since  the plasma  $\beta$ is  negligible  upstream of  the shock  and
$\beta=  0.037$ downstream.  However,  such low  values of  $\beta$ do
occur in dense molecular  clouds and protostellar discs (e.g. Crutcher
1999).  Furthermore, even though $\beta$  is small, such errors in the
gas pressure can have a significant effect on the dynamics because the
gas  pressure provides  a force  parallel  to the  field, whereas  the
Lorentz force does not.

Finally, Figure 6 shows that a  relatively small error at a fast shock
can be amplified by a slow shock following on behind. In this case the
ZEUS solution  has an error of  $22\%$ in the density  behind the slow
shock travelling  to the  right. This is  not caused by  small $\beta$
since $\beta= 0.16$ behind the fast shock, $\beta=6.1$ behind the slow
shock and  the error in the gas  pressure is much smaller  than in the
density.

Like  Balsara (2001),  we  find that  ZEUS  produces large  post-shock
oscillations  for strong  MHD shocks,  but that  these can  be reduced
substantially  by adding  the  same linear  artificial viscosity  that
removes gas  dynamic rarefaction shocks. This is  presumably because a
quadratic viscosity  leads to  algebraic decay of  these oscillations,
whereas a  linear viscosity  gives exponential decay.  The calculation
shown in Figure  6 used this value of  the linear artificial viscosity
and it can  be seen that the amplitude  of the post-shock oscillations
is quite small.

The calculations presented are all coplanar  ($v_z = B_z = 0$), but we
have also looked at some non-coplanar problems in order to see whether
the presence of Alfv\'en  waves causes any additional difficulties for
ZEUS.  This  is not the case, at  least for the problems  that we have
considered.

\section{Conclusion}

It is  evident from these  results that, ZEUS  can be made  just about
acceptable for pure gas dynamics if the linear artificial viscosity is
multiplied by the smallest local Courant number since the shock errors
are small in this case.  However, it is not satisfactory for adiabatic
MHD, at least in its present  form.  The shock errors do not occur for
an isothermal equation of state, but, since the rarefaction shocks do,
ZEUS is also not reliable for  isothermal MHD. It is possible that the
rarefaction  shocks in  MHD  waves  can be  removed  without using  an
excessive  linear   artificial  viscosity   by  the  addition   of  an
appropriate linear artificial resistivity. The shock errors might also
be  reduced by  advecting the  total energy  rather than  the internal
energy.   However,  even with  such  improvements,  the  low order  of
accuracy  makes ZEUS very  inefficient compared  with a  modern upwind
scheme.

This should not be taken to mean that conservative upwind codes are in
any sense  perfect.  For  example, it is  necessary to  introduce some
extra dissipation in  the Riemann solver to remove  the serious errors
discussed  by Quirk  (1994)  and some  desirable  properties, such  as
strict conservation, may have to be sacrificed in order to satisfy the
constraint $\nabla  \cdot {\bf  B} = 0$  in multidimensional  MHD (see
e.g. Powell et al. 1999; Balsara 2001).

These results  obviously have implications for the  reliability of the
numerous calculations in the  literature that have used ZEUS. Although
these effects are  likely to be present in  many cases, the associated
errors are not necessarily so  serious as to completely invalidate the
calculations. Whether  or not they make any  qualitative difference in
any  particular  case  can  only  be  decided  either  by  a  thorough
examination  of the results  to see  whether any  of these  errors are
present, or by repeating the  calculations using a modern code.

These calculations were performed with the version of Zeus2d available
from the NCSA  website, but, since all versions of  ZEUS appear to use
the same algorithms,  the results should not depend  on the particular
version.   It is also  worth pointing  out that  although we  used the
scheme  described by  Falle,  Komissarov and  Joarder (1998),  similar
results would probably have been obtained with any modern upwind code.

The  author would  like  to thank  both  the editor  and an  anonymous
referee for a number of helpful comments on the original version.

\end{document}